\documentclass[aps,superscriptaddress,showpacs,twocolumn]{revtex4}
\usepackage{hyperref}
\usepackage{graphicx}

\begin{document}
\title{Number of first-passage times as a measurement\\ of information for weakly chaotic systems}

\author{Pierre Naz\'e}\email{pierre.naze@ufabc.edu.br}
\address{Centro de Ci\^encias Naturais e Humanas, UFABC, 09210-170, Santo Andr\'e, SP, Brazil}
\author{Roberto Venegeroles}\email{roberto.venegeroles@ufabc.edu.br}
\address{Centro de Matem\'atica, Computa\c c\~ao e Cogni\c c\~ao, UFABC, 09210-170, Santo Andr\'e, SP, Brazil}

\date{\today}

\begin{abstract}
We consider a general class of maps of the interval having Lyapunov subexponential instability $|\delta x_{t}|\sim|\delta x_{0}|\exp[\Lambda_{t}(x_{0})\zeta(t)]$, where $\zeta(t)$ grows sublinearly as $t\rightarrow\infty$. We outline here a scheme [J. Stat. Phys. {\bf 154}, 988 (2014)] whereby the choice of a characteristic function automatically defines the map equation and corresponding growth rate $\zeta(t)$. This matching approach is based on the infinite measure property of such systems. We show that the average information that is necessary to record without ambiguity a trajectory of the system tends to $\langle\Lambda\rangle\zeta(t)$, suitably extending the Kolmogorov-Sinai entropy and Pesin's identity. For such systems, information behaves like a random variable for random initial conditions, its statistics obeying a universal Mittag-Leffler law. We show that, for individual trajectories, information can be accurately inferred by the number of first-passage times through a given turbulent phase space cell. This enables us to calculate far more efficiently Lyapunov exponents for such systems. Lastly, we also show that the usual renewal description of jumps to the turbulent cell, usually employed in the literature, does not provide the real number of entrances there. Our results are supported by exhaustive numerical simulations.
\end{abstract}


\maketitle

\section{Introduction}
\label{sec1}

After the pioneering work of Gaspard and Wang \cite{GW}, we have witnessed in recent years a growing interest in the study and characterization of dynamical systems whose separation of initially nearby trajectories is weaker than exponential. More specifically, such separation grows as \cite{GW,RV1}
\begin{equation}
\label{wc}
|\delta x_{t}|\sim|\delta x_{0}|\exp[\Lambda_{t}(x_{0})\zeta(t)],\qquad t\rightarrow\infty,
\end{equation}
with sublinear growth rate given by
\begin{equation}
\label{zetw}
\zeta(t)\sim l(t)t^{\alpha},\qquad0\leq\alpha\leq1,
\end{equation}
being $l(t)$ a slowly varying function at infinity such that $l(t\rightarrow\infty)=\infty$ for $\alpha=0$ and  $l(t\rightarrow\infty)=0$ for $\alpha=1$ \cite{note}. The coefficient $\Lambda_{t}(x_{0})$ stands for the largest finite-time Lyapunov exponent at $x_{0}$ for usual chaotic systems when $\zeta(t)\sim t$. Here it is the corresponding generalization for weakly chaotic systems that evolve according to the growth rate (\ref{zetw}).

The unpredictable nature of deterministic systems has its origin in the sensitivity to initial conditions. In this context, the degree of randomness of a system is usually characterized by the Kolmogorov-Sinai (KS) entropy. The KS entropy is the intrinsic rate at which information is produced by the dynamical system and gives the number of bits that is necessary and sufficient to record without ambiguity its trajectory during a unit time interval \cite{GW1}. For usual chaotic systems, the information $C_{t}(x)$ contained in $t$ steps of a single trajectory of a point $x$ with respect to a partition of the phase space is asymptotically related to the KS entropy $h_{\mu}$ as $C_{t}(x)\sim h_{\mu}t$, almost everywhere. The KS entropy measures the degree of unpredictability (or irregularity) of a system but not necessarily the difficulty of modeling it from experimental data \cite{RVM}. This means, of course, that more complex systems may be modeled by means of simpler ones, helping to establish a understanding of their fundamental issues.

The kind of weakly chaotic maps to be considered here have quite a few very interesting features. In particular, by considering the symbolic dynamics approach, laminar phases of dynamics correspond to repetitions of the same symbol for quite a long time, which are interrupted by successive turbulent outbursts. Thus, symbol repetitions generate long-term correlations that can not be replicated by means of usual chaotic systems. Recently, a weakly chaotic map has been employed to model the dynamics of DNA strands with long-range features, such as the genome of higher eucaryotes \cite{PB}. A similar model with external noise has also been used for modeling the time distance between consecutive neuron firings in the study of neural avalanches in the brain of mammals \cite{ZG}. Recent results also suggest that subexponential instability might be observed in some stroboscopic maps related to the FitzHugh-Nagumo model for a single externally excited neuron \cite{BS}.

\section{Preliminaries and main results}
\label{sec2}

Here we study a general class of maps of the interval, weakly chaotic in the sense of Eq. (\ref{zetw}), for which
\begin{equation}
\label{inf}
E[C_{t}(x)]\rightarrow\langle\Lambda\rangle\zeta(t), \qquad t\rightarrow\infty,
\end{equation}
where $E[\ldots]$ denotes the expected value over initial condition ensemble and $\langle\Lambda\rangle$ is a theoretical value to be established later on. Subexponential instability (\ref{zetw}) is an infinite measure property, as opposed to the usual chaotic systems, whose invariant measure is finite. Based on recent results \cite{RV1}, this relationship is quantitatively outlined here, enabling the calculation of $\zeta(t)$ and thus the estimation of average information (\ref{inf}) directly from the map equations itself. Moreover, such systems compel universal Mittag-Leffler statistics of observables \cite{Aaronson}, then our knowledge of randomness is not restricted to the first moment (\ref{inf}).

The mechanism for generating subexponential instability (\ref{zetw}) relies on the existence of two phases of motion: a laminar region where the invariant measure is infinite and, therefore, the motion is very slow, and a complementary turbulent region of short time bursts. By considering the number $N_{t}$ of first-passage times from laminar to turbulent region during $t$ iterations of the map \cite{GW}, we shall see that
\begin{equation}
\label{mrel}
C_{t}(x)\rightarrow\gamma N_{t}(x), \qquad t\rightarrow\infty
\end{equation}
almost everywhere, where $\gamma=\gamma(x_{*},\alpha)$ and $x_{*}$ defines a two-phase space partition. We will provide here a general formula for the prefactor $\gamma$ and also show that the standard partition maximizes $N_{t}$. Interestingly, we shall see that $N_{t}$ {\it is not} the number of renewals occurring at time $t$ as usually considered in the literature.

For finite measure chaotic systems, more specifically closed Anosov, the KS entropy is known to be given by the sum of positive Lyapunov exponents, the well-known Pesin identity \cite{Pes}. On the other hand, for weakly chaotic systems (\ref{zetw}), both Lyapunov exponents and KS entropy become zero if computed in the conventional manner, concealing the subtler character of their dynamic instability. Then we propose a generalization for these quantities and, accordingly, also for Pesin's relation, extending earlier investigations done in Ref. cite{SVp}.

Despite not yet being widely explored in other fields, sublinear behavior of the type (\ref{zetw}-\ref{inf}) is not restricted to infinite measure maps \cite{GW,SVp,BG}, but has also been observed in other systems such as anomalous diffusion by L\'evy flights \cite{Wang}, Feigenbaum's attractor at the threshold of chaos in the period doubling scenario \cite{Grass1}, weather systems \cite{NEB,RP}, sporadic behavior of language texts \cite{EN,Grass2}, complexity of classical integer factorization algorithms used to break RSA cryptosystems \cite{Pome} as well as of some noncoding DNA sequences \cite{EFH,LK}.

\section{Maps with sublinear growth rate}
\label{sec4}

Let us now introduce the general class of piecewise $C^{2}$ expanding maps $x_{t+1}=T(x_{t})$, from $[0,1]$ to itself, such that $T:(0,c)\rightarrow(0,1)$ and $T:(c,1)\rightarrow(0,1)$. On the interval $(0,c)$ the map $T$ is given by
\begin{equation}
\label{map}
 T(x)=x+f(x),
\end{equation}
with a single marginal fixed point at $x=0$, i.e.,
\begin{equation}
\label{marg}
 f(x\rightarrow 0)=f'(x\rightarrow 0)=0.
\end{equation}
Since these conditions are met, the overall shape of $T$ far from the marginal fixed point is of secondary importance (a sketch of $T$ will be shown later in Sec. \ref{sec7}). The best known models that meet these criteria are the Pomeau-Manneville (PM) type of maps
\begin{equation}
\label{PM}
T(x)\sim x+ax^{1+1/\alpha},\qquad x\rightarrow0,
\end{equation}
where $a$ and $\alpha$ are positive parameters. For $\alpha=1$ one has the original PM map, obtained from Poincar\'e sections related to the Lorenz attractor \cite{PM}.

The weakly chaotic behavior (\ref{zetw}) stems from the divergence of invariant measure $d\mu(x)=\rho(x)dx$ near the marginal fixed point. For example, for the case of PM map (\ref{PM}) one has \cite{Thaler}
\begin{equation}
\label{rhopm}
\rho(x)\sim bx^{-1/\alpha},\qquad x\rightarrow0,
\end{equation}
and thus weak chaos for $0<\alpha\leq1$. It is important to emphasize that, for infinite measure systems, $\rho(x)$ is defined up to a positive multiplicative constant since there is no possible normalization. In other words, $\mu$ is not a probability measure for such systems. Therefore, the value of $b$ in Eq. (\ref{rhopm}) is {\it unreachable} for $0<\alpha\leq1$ \cite{SVp,SVpr}.

According to recent results \cite{RV1}, the spatial and temporal properties of systems like (\ref{map}-\ref{marg}) are determined by the simple choice of a characteristic function $\phi(x)$ satisfying
\begin{equation}
\label{infp}
\int\phi(x)dx\rightarrow-\infty,\qquad x\rightarrow0.
\end{equation}
The invariant density $\rho(x)$ gives the measure of concentration trajectories on phase space, whereas the residence times in the laminar region are ruled by the waiting-time density function $\psi(t)$. According to \cite{RV1} one has
\begin{equation}
\label{frhop}
f'(x)\sim\frac{1}{\rho(x)}\sim-\frac{1}{x\phi'(x)},\qquad x\rightarrow0,
\end{equation}
up to positive multiplicative constants, and
\begin{equation}
\label{psiv}
\psi(t)\sim-[\phi^{-1}(t)]',\qquad t\rightarrow\infty,
\end{equation}
$\phi^{-1}$ denoting the inverse of $\phi$. Moreover, the Laplace transform $\tilde{\zeta}(s)$ of growth rate is
\begin{equation}
\label{zetal}
\tilde{\zeta}(s)\sim\frac{1}{s[1-\tilde{\psi}(s)]},\qquad s\rightarrow0.
\end{equation}

The quantitative connection of $\zeta(t)$ with the infinite invariant measure was recently shown in \cite{RV1}: the divergence of mean waiting time $\int_{0}^{\infty}t\psi(t)dt$ implies infinite invariant measure near the marginal fixed point. By considering the general form of cumulative distribution function associated to $\psi(t)$, i.e., $\int_{0}^{t}\psi(u)du\sim 1-1/q(t)t^{\alpha}$, $q(t)$ being a slowly varying function at infinity and $0\leq\alpha\leq1$, one has
\begin{eqnarray}
\label{psiexp}
\tilde{\psi}(s) \sim \left\{
\begin{array}{ll}
  1-\Gamma(1-\alpha)s^{\alpha}/|q(1/s)|,  & 0 \leq \alpha < \displaystyle 1, \\
  1-s|\ln s|/|q(1/s)|, & \displaystyle \alpha=1,
\end{array}
\right.
\end{eqnarray}
as $s\rightarrow0$. Now, noting also that $\int_{0}^{t}\psi(u)du\sim 1-\phi^{-1}(t)$ from Eq. (\ref{psiv}), the growth rate formula (\ref{zetal}) can be solved by using Karamata's Abelian and Tauberian theorems for the Laplace-Stieltjes transform \cite{Feller}, resulting
\begin{eqnarray}
\label{zp}
\zeta(t)\sim \frac{1}{\phi^{-1}(t)}\times\left\{
\begin{array}{ll}
  \sin(\pi\alpha)/\pi\alpha,  & 0\leq\alpha < \displaystyle 1, \\
  1/\ln t, & \displaystyle \alpha=1.
\end{array}
\right.
\end{eqnarray}

The simple choice of $\phi(x)$ near the marginal fixed point satisfying condition (\ref{infp}) is sufficient to construct a variety of ergodicly equivalent maps. Let us now consider PM type of maps near $x=0$ in the weakly chaotic regime $0<\alpha\leq1$ (other possibilities of maps are discussed in \cite{RV1}). Such maps are given by the choice
\begin{equation}
\label{pm}
\phi(x)\sim\frac{\alpha}{a}x^{-1/\alpha},\qquad x\rightarrow0,
\end{equation}
for which $f(x)\sim-1/\phi'(x)$. The corresponding invariant density evaluated by Eq. (\ref{frhop}) results on Eq. (\ref{rhopm}). For $0<\alpha<1$ we have from Eq. (\ref{zp})
\begin{equation}
\label{tab}
\zeta(t)\sim\left(\frac{a}{\alpha}\right)^{\alpha}\frac{\sin(\pi\alpha)}{\pi\alpha}t^{\alpha},
\end{equation}
whereas for $\alpha=1$
\begin{equation}
\label{tab1}
\zeta(t)\sim a\frac{t}{\ln t}.
\end{equation}

\section{Distributional convergence and Pesin-type relation}
\label{sec6}

For a one-dimensional chaotic map $T$ with finite measure $\mu$ absolutely continuous with respect to the Lebesgue measure, Pesin's relation implies that $\Lambda_{t}(x)\rightarrow h_{\mu}$ as $t\rightarrow\infty$, and thus leading to
\begin{equation}
\label{bp}
C_{t}(x)\rightarrow\sum_{k=0}^{t-1}\ln|T'[T^{k}(x)]|,\qquad t\rightarrow\infty
\end{equation}
almost everywhere. Furthermore, according to the well-known Birkhoff theorem, ergodicity implies $t^{-1}C_{t}(x)\rightarrow h_{\mu}=\int\ln|T'|d\mu$.

For infinite measure systems Eq. (\ref{bp}) still holds \cite{Zei,RVT}, but within a completely different scenario: despite the pointwise convergence, different initial conditions lead to different values of $C_{t}(x)$ and $\Lambda_{t}(x)$ no matter how long the iteration time. In other words, even taking into account a suitable time average of an observable, it does not converge to a constant value, actually behaving like a random variable for random initial conditions. To be more specific, for a positive integrable function $\vartheta$ and a random variable $x$ with an absolutely continuous measure with respect to the Lebesgue measure, the Aaronson-Darling-Kac convergence in distribution takes place \cite{Aaronson}:
\begin{equation}
\label{adk}
\frac{1}{a_{t}}\frac{\sum_{k=0}^{t-1}\vartheta[T^{k}(x)]}{\int\vartheta d\mu}\stackrel{d}{\longrightarrow}\xi_{\alpha},\qquad t\rightarrow\infty,
\end{equation}
where $a_{t}$ is the so-called return sequence and $\xi_{\alpha}$ is a Mittag-Leffler random variable with index $\alpha\in[0,1)$ and unit expected value. The corresponding Mittag-Leffler probability density function is \cite{SVa}
\begin{equation}
\label{mldens}
p_{\alpha}(\xi)=\frac{\Gamma^{1/\alpha}(1+\alpha)}{\alpha\xi^{1+1/\alpha}}\,g_{\alpha}\left[\frac{\Gamma^{1/\alpha}(1+\alpha)}{\xi^{1/\alpha}}\right],
\end{equation}
being $g_{\alpha}(u)$ the one-sided L\'evy density function, i.e., $\tilde{g}_{\alpha}(s)=\exp(-s^{\alpha})$. As $\alpha\rightarrow1$, the width of Mittag-Leffler density tends to zero, resembling an approximation of $\delta(\xi-1)$, as is typical of finite measure ergodic systems. As $\alpha\rightarrow0$, Eq. (\ref{mldens}) reads the exponential density $p_{0}(\xi)=e^{-\xi}$.

The results developed in Ref. \cite{RV1} also provide a direct way to obtain return sequences for such systems, namely
\begin{equation}
\label{rszeta}
a_{t}=\kappa\zeta(t),
\end{equation}
being $\kappa$ given by
\begin{equation}
\label{kap}
\kappa=-\lim_{x\rightarrow0}\frac{x\phi'(x)}{\rho(x)}.
\end{equation}
For PM maps we have
\begin{equation}
\label{psit0}
\kappa=\frac{1}{ab}
\end{equation}
for $0<\alpha\leq1$. Note that $\kappa\zeta(t)$ given by Eqs. (\ref{tab}),  (\ref{tab1}), and (\ref{psit0}) are identical to $a_{t}$ according to the infinite ergodic theory \cite{TZ}.

From Eq. (\ref{wc}), the finite-time (generalized) Lyapunov exponent of map $T$ is
\begin{equation}
\label{laver}
\Lambda_{t}(x)=\frac{1}{\zeta(t)}\sum_{k=0}^{t-1}\ln|T'[T^{k}(x)]|,
\end{equation}
and the choice $\vartheta=\ln|T'|$ gives us (see also Ref. \cite{PSV})
\begin{equation}
\label{Ladk}
\frac{\Lambda_{t}(x)}{\langle\Lambda\rangle}\stackrel{d}{\longrightarrow}\xi_{\alpha},\qquad t\rightarrow\infty,
\end{equation}
with
\begin{equation}
\label{Lav}
\langle\Lambda\rangle=\kappa\int\ln|T'|d\mu.
\end{equation}
It is important to emphasize here that the unreachable constant prefactor of invariant density cancels itself out in Eq.  (\ref{Lav}). Moreover, $\ln|T'|\sim f'=O(1/\rho)$ near $x=0$, and thus $\langle\Lambda\rangle$ is always a well defined quantity.

Based on Eq. (\ref{bp}), the natural generalization of Pesin's identity is as follows
\begin{equation}
\label{preal}
\frac{C_{t}(x)}{\zeta(t)}\equiv h_{t}(x)\rightarrow\Lambda_{t}(x),\qquad t\rightarrow\infty,
\end{equation}
being $h_{t}(x)$ the (generalized) finite-time KS entropy. Relationship (\ref{preal}) and the general growth rate formula (\ref{zp}) encompass Pesin's formula proposed in Ref. \cite{SVp}. Moreover, we have
\begin{equation}
\label{hml}
\frac{C_{t}(x)}{\langle \Lambda\rangle\zeta(t)}\stackrel{d}{\longrightarrow}\xi_{\alpha},\qquad t\rightarrow\infty.
\end{equation}
Thus, we have not only the average information previously introduced in Eq. (\ref{inf}), which enables us to record the behavior of trajectories, but also its statistical law.

\section{First-passage times through the turbulent phase space cell}
\label{sec7}

Let us now consider the usual partition of the interval $[0,1]$ into two cells, $A_{0}=[0,x_{*}]$ and $A_{1}=(x_{*},1]$. According to the symbolic dynamics approach, a given trajectory $\{x_{t}\}$ generated by $T$ can be represented by a sequence of integer symbols $\{s_{t}\}$ such that $s_{t}$ corresponds to the cell where $x_{t}$ belongs, for example, $s_{t}=k$ for $x_{t}\in A_{k}$. Then we eliminate the redundancies that may appear in $\{s_{t}\}$ by performing a compression of information. This is accomplished by considering the algorithmic information $C_{t}$, which is defined as the length of the shortest possible program able to reconstruct the sequence $\{s_{t}\}$ on a universal machine \cite{GW}.

\begin{figure}[!htb]
\centering
\includegraphics[scale=0.7]{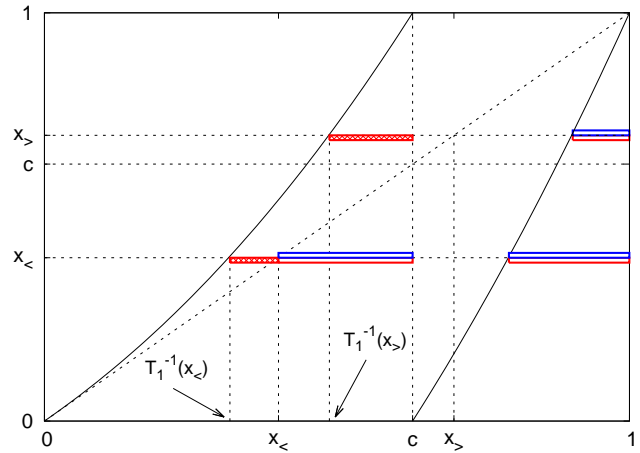}
\caption{(Color online) Schematic calculation of $\int\chi d\mu$ for a typical two branches map. For the horizontal line starting from $x_{<}$ ($x_{>}$), blue upper and red lower stripes denote intervals for which $\sigma(x)\sigma[T(x)]$ and $\sigma[T(x)]$ have nonzero measurements, respectively. Crosshatched stripes depict differences between corresponding intervals, i.e., nonzero measurement of $\chi$ according to Eq. (\ref{fpc}).}
\label{fig1}
\end{figure}

For the case of PM maps, $C_{t}(x)$ has been considered as proportional to the number of first-passage times $N_{t}(x)$ through the cell $A_{1}$ up to time $t$ \cite{GW}. Recent numerical analysis have shown that the proportionality coefficient depends on $x_{*}$ and $\alpha$ once it has fixed further map parameters \cite{SVp}. In order to understand precisely how these quantities are related, let us introduce the following observable
\begin{equation}
\label{fpc}
\chi(x)=\sigma[T(x)]-\sigma(x)\sigma[T(x)],
\end{equation}
where the indicator function $\sigma(x)$ is defined as
\begin{eqnarray}
\label{sig}
\sigma(x)=\left\{
\begin{array}{ll}
  1,  & \displaystyle x\in A_{1}, \\
  0, & \displaystyle x\in A_{0}.
\end{array}
\right.
\end{eqnarray}
Filter (\ref{fpc}) returns $1$ whenever a transition from $A_{0}$ to $A_{1}$ arises, and $0$ otherwise. By Stepanov-Hopf ratio ergodic theorem \cite{RWsh}
\begin{equation}
\label{sherg}
\frac{\sum_{k=0}^{t-1}\vartheta[T^{k}(x)]}{\sum_{k=0}^{t-1}\varphi[T^{k}(x)]}\rightarrow\frac{\int\vartheta d\mu}{\int\varphi d\mu}
\end{equation}
holds almost everywhere as $t\rightarrow\infty$ for non-negative observables (since it is integrable over $\mu$). Then we can choose $\vartheta=\ln|T'|$ and $\varphi=\chi$, resulting $N_{t}(x)=\sum_{k=0}^{t-1}\chi[T^{k}(x)]$ and relationship (\ref{mrel}), being $\gamma$ given by
\begin{equation}
\label{gnew}
\gamma=\frac{\langle\Lambda\rangle}{\kappa\int\chi d\mu}.
\end{equation}
Calculation of $\int\chi d\mu$ follows from Fig. \ref{fig1}, resulting
\begin{eqnarray}
\label{sig}
\int\chi d\mu=\left\{
\begin{array}{ll}
  \mu[T_{1}^{-1}(x_{*}),x_{*}],  & \displaystyle x_{*}\leq c, \\
  \mu[T_{1}^{-1}(x_{*}),c], & \displaystyle x_{*}>c.
\end{array}
\right.
\end{eqnarray}

Quantity (\ref{sig}) has a unique global maximum at $x_{*}=c$ (standard partition). To prove this it suffices to consider the variation $x_{*}\mapsto x_{*}+\delta x_{*}$ and the Perron-Frobenius formula \cite{RV1} to get
\begin{equation}
\label{insp1}
\delta\int\chi d\mu=(-1)^{k}\frac{\rho[T_{k}^{-1}(x_{*})]}{T'[T_{k}^{-1}(x_{*})]}\delta x_{*},
\end{equation}
where $k=1$ for $x_{*}>c$ and $k=2$ for $x_{*}\leq c$ denote the two expanding branches of $T$. Therefore, the slope $\gamma$ has a unique global minimum at $x_{*}=c$, thereby maximizing the number of entrances $N_{t}$.

Lastly, distributional formula  (\ref{adk}) applied to $\chi$ yields
\begin{equation}
\label{Nfpc}
\frac{N_{t}(x)}{\langle N_{t}\rangle}\stackrel{d}{\longrightarrow}\xi_{\alpha},\qquad \langle N_{t}\rangle=\kappa\zeta(t)\int\chi d\mu.
\end{equation}
It is also noteworthy that our result (\ref{mrel}), together with Eqs. (\ref{zp}) and (\ref{preal}), enables us to calculate far more efficiently Lyapunov exponents (\ref{laver}), i.e., for almost all $x$ one has
\begin{equation}
\Lambda_{t}(x)\rightarrow\gamma\frac{N_{t}(x)}{\zeta(t)},\qquad t\rightarrow\infty.
\end{equation}
It is much more simple to calculate the number of entrances $N_{t}$ than the summation $\sum_{k=0}^{t-1}\ln|T'[T^{k}(x)]|$. For this, the knowledge $\gamma$ is crucial.

\section{Failure of renewal description}
\label{sec8}

In many studies on the (infinite) ergodic properties of PM maps, the number of first-passage times $N_{t}$ has been usually considered to be the number of renewals in the time interval $(0,t)$, see Refs. \cite{GW,KB,RVT,Wpra}, among others. According to this approach, the sequences of positive waiting times $t_{1}, t_{2}, t_{3}, \ldots$ are considered as independent identically distributed random variables, with probability density $\psi(t)\sim t^{-(1+\alpha)}$. Therefore, for $0<\alpha\leq1$, one has infinite mean waiting time in the laminar state $A_{0}$, at full contrast with the turbulent state $A_{1}$, having a characteristic average time. Under these assumptions, the statistics of the number of transitions $N_{t}$ between the two states up to time $t$ is such that \cite{Feller}
\begin{equation}
\label{nren}
\tilde{p}_{N_{t}}(s)=\tilde{\psi}^{N_{t}}(s)\frac{[1-\tilde{\psi}(s)]}{s}.
\end{equation}

For $0\leq\alpha<1$, Eq. (\ref{psiexp}) gives us
\begin{equation}
\label{ntruc}
\tilde{p}_{N_{t}}(s)\rightarrow-\frac{d}{dN_{t}}\left\{\frac{1}{s}\exp[-N_{t}\tilde{r}_{\alpha}(s)s^{\alpha}]\right\},
\end{equation}
where $\tilde{r}_{\alpha}(s)=\Gamma(1-\alpha)/|q(1/s)|$. By taking the inverse of Laplace transform we get
\begin{equation}
\label{pnts}
p_{N_{t}}(t)\rightarrow-\frac{d}{dN_{t}}\int_{0}^{t}\frac{1}{[N_{t}r_{\alpha}(t)]^{1/\alpha}}\,g_{\alpha}\left\{\frac{u}{[N_{t}r_{\alpha}(t)]^{1/\alpha}}\right\}du,
\end{equation}
provided the following condition is fulfilled:
\begin{equation}
\label{cond}
\frac{tr'_{\alpha}(t)}{r_{\alpha}(t)}=-\frac{tq'(t)}{q(t)}\rightarrow0.
\end{equation}
A function $q$ is slowly varying if and only if there exists $\tau_{0}>0$ such that for all $\tau\geq\tau_{0}$ the function can be written in the form \cite{GS}
\begin{equation}
q(t)=\exp\left[\eta(t)+\int_{\tau_{0}}^{t}\frac{\epsilon(\tau)}{\tau}d\tau\right],
\end{equation}
where $\eta(t)\rightarrow\mbox{const.}$ and $\epsilon(t)\rightarrow0$ as $t\rightarrow\infty$. Thus condition (\ref{cond}) is fulfilled because $t\eta'(t)\rightarrow0$. Finally, solving Eq. (\ref{pnts}) one obtains
\begin{equation}
\label{pntm}
{p}_{N_{t}}(t)\rightarrow\frac{t}{\alpha N_{t}^{1+1/\alpha}[r_{\alpha}(t)]^{1/\alpha}}\,g_{\alpha}\left\{\frac{t}{[N_{t}r_{\alpha}(t)]^{1/\alpha}}\right\},
\end{equation}
and the simple change of variable $d\xi=dN_{t}/\zeta(t)$ leads $p_{N_{t}}(t)$ to Eq. (\ref{mldens}) where
\begin{equation}
\label{Nadk}
\frac{N_{t}(x)}{\langle N_{t}\rangle}\stackrel{d}{\longrightarrow}\xi_{\alpha},\qquad \langle N_{t}\rangle=\zeta(t).
\end{equation}

Thus, we can see that the renewal description of intermittent jumps is, at best, a qualitative description of this phenomenon: although Mittag-Leffler statistics remains asymptotically valid, the average number of renewals (\ref{Nadk}) does not match with the real number of entries in $A_{1}$, given by Eq. (\ref{Nfpc}), namely $\langle N_{t}\rangle/\zeta(t)=\langle\Lambda\rangle/\gamma$. For example, for $\alpha=1/2$ and standard partition, Thaler's map [Eq. (\ref{thmap}) below] gives $\langle\Lambda\rangle/\gamma=\sqrt{2}-1$, showing that approximately $59\%$ of the counts in the renewal description are theoretically overestimated. Results that depend qualitatively on the statistics of ratio $N_{t}(x)/\langle N_{t}\rangle$ are not affected when considering the renewal approach, as in Ref. \cite{GW} or in the thermodynamic formalism of PM maps \cite{RVT,Wpra}. However, it is very important to notice that the number of entrances (\ref{Nadk}) does not depend on the partition of phase space, whereas $\langle\Lambda\rangle/\gamma$ becomes increasingly smaller as it moves away from standard partition (see Fig. (\ref{fig3}) below). Thus, the algorithm complexity $C_{t}$ can not be taken as proportional to the number of renewals in the sense of Eq. (\ref{Nadk}).

\section{Numerical simulations}
\label{sec9}

In order to check and illustrate our main results, we perform an exhaustive numerical analysis by considering the Thaler map $T:[0,1]\rightarrow[0,1]$ defined by \cite{Thaler}
\begin{equation}
\label{thmap}
T(x)=x\left[1+\left(\frac{x}{1+x}\right)^{\frac{1-\alpha}{\alpha}}-x^{\frac{1-\alpha}{\alpha}}\right]^{-\frac{\alpha}{1-\alpha}}\,\,\,\mbox{mod}\,1.
\end{equation}
This map is very useful for our purposes because its invariant density is explicitly known throughout the interval, namely \cite{Thaler}
\begin{equation}
\label{thinvd}
\rho(x)=b[x^{-\frac{1}{\alpha}}+(1+x)^{-\frac{1}{\alpha}}],
\end{equation}
where $b$ is the undefined constant for $0<\alpha\leq1$. Moreover, its behavior near $x=0$ is the same of PM map (\ref{PM}) for $a=1$, thus we can also use directly the auxiliary results of Sec. \ref{sec4}.

\begin{figure}[!b]
\centering
\includegraphics[scale=0.7]{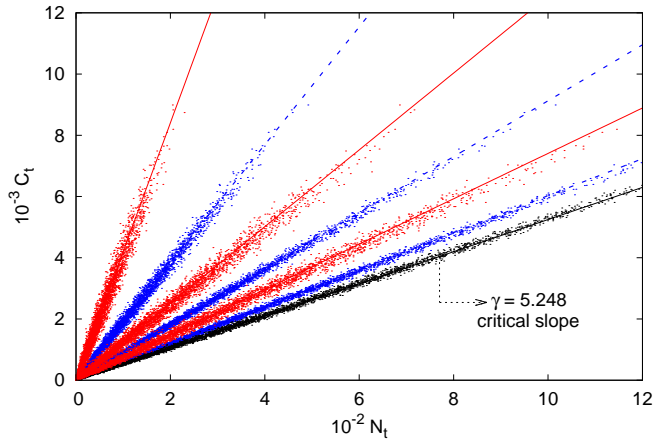}
\caption{(Color online) Graphics of the algorithm complexity $C_{t}$, calculated by Eq. (\ref{bp}), as a function of $N_{t}$, the number of entrances of a given trajectory into the cell $A_{1} = (x_{*},1]$ during $t=10^6$ iterations of Thaler's map with $\alpha=1/2$. The slopes $\gamma$ for each straight line were calculated according to Eqs. (\ref{gnew}) and (\ref{sig}) for the same map. Each straight line is matched by $4\times10^{3}$ numerical points obtained from initial conditions uniformly distributed on the interval $[0,1]$. Solid red (dashed blue) lines depict partitions with $x_{*}>c$ ($x_{*}<c$), $x_{*}$ increasing in the counterclockwise (clockwise) direction. Critical line comes from the standard partition $x_{*}=c$. In all cases, the error in $\gamma$ is less than $2\%$ of the corresponding theoretical value.}
\label{fig2}
\end{figure}

\begin{figure}[!b]
\centering
\includegraphics[scale=0.7]{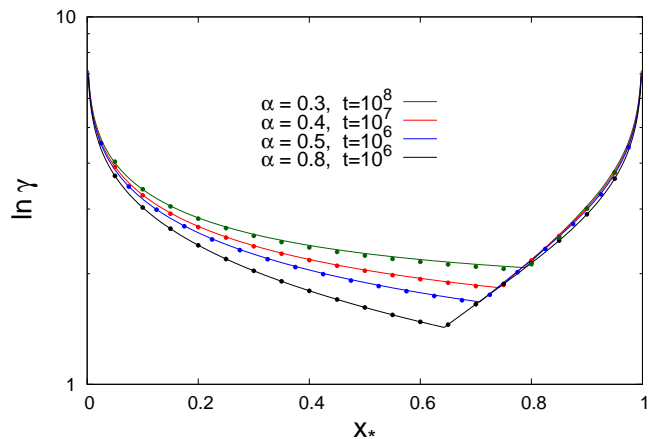}
\caption{(Color online) Solid lines from top ($\alpha=0.3$) to bottom ($\alpha=0.8$) represent the semi-log plot of $\gamma$ according to Eqs. (\ref{gnew}) and (\ref{sig}) for Thaler's map. Each numerical point was obtained from the map iterations and by least squares fitting of $C_{t}\rightarrow\gamma N_{t}$ over $10^{3}$ initial conditions uniformly distributed on the interval $[0,1]$.}
\label{fig3}
\end{figure}

First, Fig. (\ref{fig2}) depicts relationship (\ref{mrel}). As one can see, both quantities are indeed proportional, irrespective of the partition employed. The choice of partition determines the slope of straight lines, which are in full agreement with Eqs. (\ref{gnew}) and  (\ref{sig}). The standard partition $x_{*}=c$ leads to a global minimal slope, also in full agreement with our predictions in Sec. (\ref{sec7}). Figure (\ref{fig3}) depicts the dependence of slope $\gamma$ with the partition of phase space, explicitly highlighting the global minimum at $x_{*}=c$. Note that small values of $\alpha$ require larger simulation times since, in such cases, trajectories spend much more time in the laminar region.

\begin{figure}[!h]
\centering
\includegraphics[scale=0.7]{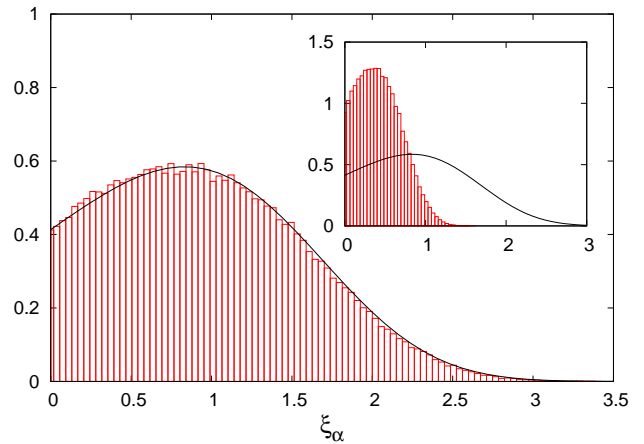}
\caption{(Color online) Distribution of the random variable $\xi_{\alpha}=N_{t}/\langle N_{t}\rangle$ for Thaler's map with $\alpha=2/3$, $t=10^{8}$, and standard partition of phase space. Both histograms were built from the same $2.5\times10^{5}$ initial conditions uniformly distributed on the interval $[0,1]$. The background and the inset plots correspond, respectively, to rates $\xi_{\alpha}$ whereby $\langle N_{t}\rangle$ was employed according to Eqs. (\ref{Nfpc}) and (\ref{Nadk}). For both cases, solid line is the Mittag-Leffler probability density with $\alpha=2/3$ and unit expected value. As one can see, the background distribution behaves according to our prediction in Sec. \ref{sec7}, contrary to the renewal approach provides (inset).}
\label{fig4}
\end{figure}

Lastly, we can check the distributional limit (\ref{Nfpc}), i.e., if the normalized ratio $N_{t}(x)/\langle N_{t}\rangle$ does converge toward a Mittag-Leffler distribution with unit expected value. Figure (\ref{fig4}) shows us an example for the Thaler map where both histograms were built directly from the same numerical data, one being normalized with $\langle N_{t}\rangle=52593.6$, according to Eq. (\ref{Nfpc}), the other by the renewal average given in Eq. (\ref{Nadk}), namely $\langle N_{t}\rangle=116734.6$. The corresponding numerical value is $\langle N_{t}\rangle=52591.9$. The agreement with results of Sec. \ref{sec7} and the failure of renewal approach are evident.

\section{Other Pesin-type relations}
\label{sec10}

The idea of extending Pesin's relation for subexponential systems has been quite contentious in recent years. Our Pesin's relation (\ref{preal}) is an extension of the relation recently proposed in Ref. \cite{SVp} for PM maps, which can be confirmed by a simple inspection via Eqs. (\ref{tab}) and (\ref{laver}).

A few years before, Barkai and Korabel (BK) proposed a different Pesin-type formula for PM maps in \cite{KB}. They considered the quantity
\begin{equation}
\label{bklyp}
\frac{1}{t^{\alpha}}\sum_{k=0}^{t-1}\ln|T'[T^{k}(x)]|\rightarrow\lambda_{\alpha}(x)
\end{equation}
as the generalized Lyapunov exponent for such systems. Their Pesin-type formula for PM maps is as follows
\begin{equation}
\label{pdle}
\mbox{BK relation:}\,\,\, h_{\alpha}=\alpha\langle \lambda_{\alpha}\rangle,
\end{equation}
where $h_{\alpha}=\int\ln|T'(x)|\rho(x)dx$ is Krengel's entropy.

In a very recent paper \cite{BK2}, BK claim that the average over initial conditions of Pesin's relation proposed in Ref.  \cite{SVp} is {\it exactly} their previously obtained result, i.e., Eq. (\ref{pdle}). Pesin's relation proposed in Ref. \cite{SVp} is equivalent to our relation (\ref{preal}) for PM maps, and we can show that its average over initial conditions {\it does not coincide} with the BK relation (\ref{pdle}). By taking the average of Eq. (\ref{mrel}) over initial conditions, our results in Sec. \ref{sec7} lead simply to 
\begin{equation}
\label{avov}
E\left[\frac{C_{t}(x)}{\zeta(t)}\right]\rightarrow\gamma E\left[\frac{N_{t}(x)}{\zeta(t)}\right]\rightarrow\kappa h_{\alpha}.
\end{equation}
Now, by considering Eqs. (\ref{tab}) and (\ref{psit0}) for $0<\alpha<1$, and also Eq. (\ref{bp}), we finally get
\begin{equation}
\label{pour}
E[\lambda_{\alpha}(x)]\equiv\langle \lambda_{\alpha}\rangle=\left[\frac{1}{a}\left(\frac{a}{\alpha}\right)^{\alpha}\frac{\sin(\pi\alpha)}{\pi\alpha}\right]\frac{h_{\alpha}}{b},
\end{equation}
again corroborating similar results of Refs. \cite{SVp} and \cite{SVpr}.

Obviously, $b$ cancels itself out in Eq. (\ref{pour}) because $h_{\alpha}$ is proportional to $b$, see Eq. (\ref{rhopm}) or Thaler's invariant density (\ref{thinvd}). In contrast, BK relation (\ref{pdle}) depends on $b$ via $h_{\alpha}$. The average value $\langle \lambda_{\alpha}\rangle$ can always be inferred by means of numerical simulations, but the prefactor $b$ of $h_{\alpha}$  is {\it unreachable} because the invariant density is non-normalizable, see our discussion in Sec. \ref{sec4}. Thus, BK relation (\ref{pdle}) can never match Eq. (\ref{pour}), both are completely different from one another.

Note further that the use of renewal approach in Ref. \cite{KB} gives us an additional reason to conclude that the BK relation (\ref{pdle}) can never be equal to the result (\ref{pour}). Renewal approach leads to $E[N_{t}(x)/\zeta(t)]\rightarrow1$, as we have demonstrated in Sec. \ref{sec8}, whereas $E[N_{t}(x)/\zeta(t)]\rightarrow\kappa\int\chi d\mu$ according to our results in Sec. \ref{sec7}. Note also that these two limits, together with $\gamma$, do not depend on $b$. Thus, once again, Eq. (\ref{avov}) can not yield the results (\ref{pdle}) and (\ref{pour}) simultaneously. It is also noteworthy that our Eq.  (\ref{avov}) is in full agreement with numerical simulations of Sec. \ref{sec9}.

We close by noting that  Eq. (\ref{pour}) was numerically confirmed for the Thaler map (\ref{thmap}) by means of the quantity $R(\alpha)\equiv\langle \lambda_{\alpha}\rangle/(h_{\alpha}/b)$, see Ref. \cite{SVpr}.

\section{Concluding remarks}
\label{sec11}

Characterization of information is an important issue for a deep understanding of dynamical systems. We have shown that sublinear growth rate $\zeta(t)$ plays a direct role in the predictability of subexponential unstable systems. First we outline recent results relating $\zeta(t)$ with the behavior of map equations near the marginal fixed point \cite{RV1}, and then we provide a complete description of universal statistical law of algorithmic complexity for such systems. We demonstrate the linear relation between algorithmic complexity and the number of first-passage times through a given turbulent phase space cell, originally proposed in Ref. \cite{GW}. In particular, we provide a general formula for the ratio between these two quantities, enabling the computation of algorithmic complexity and (generalized) Lyapunov exponents not only accurately, but also in a much more efficient manner. Last but not least, we show that the usual renewal description of jumps to a turbulent cell does not provide the real number of entrances there. A correct formulation for the problem was also provided here by means of the results of Sec. \ref{sec7}. We also correct misleading comparisons that have appeared recently in the literature between Pesin's relation for subexponential systems and an alternative relation based on Krengel's entropy.

Despite the main features of infinite measure maps being ruled by the local singular behavior of invariant measure, certain nonlocal aspects of information demand a more general knowledge of the measurement. A full characterization of the measure is central in ergodic theory because it is the invariant measure that characterizes the occupation probabilities over entire phase space. Methods that enable complete determination of invariant densities are therefore in demand.

Much of what has been discussed here is not restricted to infinite measure maps. For example, our results are, in essence, also valid for PM type of systems with $\alpha>1$, when the invariant measure is a probability measure, growth rates are linear over time, and Mittag-Leffler density becomes a Dirac $\delta$ function, which is typical of finite measure ergodic systems. Thus, we believe that the idea of being able to estimate information by observing the dynamics in key regions of phase space is still an open field for investigations even for other types of dynamical systems \cite{RV}. We hope that the present study also brings further perspectives on the subject matter.

\begin{acknowledgements}
The authors thank Alberto Saa and Carlos J.A. Pires for helpful discussions. We acknowledge financial support by Funda\c c\~ao Universidade Federal do ABC (UFABC), Brazil, and Coordena\c c\~ao de Aperfei\c coamento de Pessoal de N\'ivel Superior (CAPES), Brazil. R.V. was also supported by Conselho Nacional de Desenvolvimento Cient\'ifico e Tecnol\'ogico (CNPq), Brazil (Grant No. 307618/2012-9), Funda\c c\~ao de Amparo \`a Pesquisa do Estado de S\~ao Paulo (FAPESP), Brazil (Grant No. 2013/03990-1), and special program PROPES-Multicentro (UFABC), Brazil.
\end{acknowledgements}


\begin{thebibliography}{99}
\bibitem{note} It is noteworthy here that slowly varying functions in Eq. (\ref{zetw}) are not restricted to powers of logarithms as in the similar expression originally conjectured in Ref. \cite{GW}, see Ref. \cite{RV1} for more details.
\end{thebibliography}
\end{document}